\documentclass[conference]{IEEEtran}
\IEEEoverridecommandlockouts
\usepackage{soul}
\usepackage{xcolor,colortbl}
\usepackage{subcaption}
\usepackage{xspace}

\usepackage[english]{babel}

\usepackage{rotating}
\usepackage{tabularx}
\usepackage{amsmath}
\usepackage{comment}
\usepackage{listings}
\usepackage{hyperref}

\definecolor{codegreen}{rgb}{0,0.6,0}
\definecolor{codegray}{rgb}{0.5,0.5,0.5}
\definecolor{codepurple}{rgb}{0.58,0,0.82}
\definecolor{backcolour}{rgb}{0.95,0.95,0.92}
\definecolor{commentsColor}{rgb}{0.497495, 0.497587, 0.497464}
\definecolor{keywordsColor}{rgb}{0.000000, 0.000000, 0.635294}
\definecolor{stringColor}{rgb}{0.558215, 0.000000, 0.135316}
\lstdefinestyle{mystyle}{
  backgroundcolor=\color{backcolour},   
  basicstyle=\footnotesize,        
  breakatwhitespace=false,         
  breaklines=true,                 
  captionpos=b,                    
  commentstyle=\color{commentsColor}\textit,    
  deletekeywords={...},            
  escapeinside={\%*}{*)},          
  extendedchars=true,              
  frame=tb,                        
  keepspaces=true,                 
  keywordstyle=\color{keywordsColor}\bfseries,       
  language=Python,                 
  otherkeywords={*,...},           
  numbers=left,                    
  numbersep=5pt,                   
  numberstyle=\tiny\color{commentsColor}, 
  rulecolor=\color{black},         
  showspaces=false,                
  showstringspaces=false,          
  showtabs=false,                  
  stepnumber=1,                    
  stringstyle=\color{stringColor}, 
  tabsize=2,                     
  title=\lstname,                  
  columns=fixed                    
}
\lstdefinelanguage{YARA}{
  keywords={rule, meta, strings, condition, matches, rules, externals},
  keywordstyle=\color{blue}\bfseries,
  ndkeywords={and, match, callback},
  ndkeywordstyle=\color{darkgray}\bfseries,
  identifierstyle=\color{black},
  sensitive=false,
  comment=[l]{//},
  morecomment=[s]{/*}{*/},
  commentstyle=\color{purple}\ttfamily,
  stringstyle=\color{red}\ttfamily,
  morestring=[b]',
  morestring=[b]"
}
\def\BibTeX{{\rm B\kern-.05em{\sc i\kern-.025em b}\kern-.08em
    T\kern-.1667em\lower.7ex\hbox{E}\kern-.125emX}}

\newcommand{\G}{\mathcal{G}\xspace}
\newcommand{\V}{\mathcal{V}\xspace}
\newcommand{\E}{\mathcal{E}\xspace}
\newcommand{\lb}{\mathcal{L}\xspace}
\newcommand{\pr}{\mathcal{P}\xspace}
\newcommand{\RoE}{\mathcal{R}\xspace}
\newcommand{\R}{\mathcal{R}\xspace}

\begin{document}

\title{IRSKG: Unified Intrusion Response System Knowledge Graph Ontology for Cyber Defense}



\author{\IEEEauthorblockN{Damodar Panigrahi\IEEEauthorrefmark{1},
Shaswata Mitra\IEEEauthorrefmark{2}, 
Subash Neupane\IEEEauthorrefmark{3},
Sudip Mittal\IEEEauthorrefmark{4},
Benjamin A. Blakely\IEEEauthorrefmark{5},
}
\IEEEauthorblockA{Dept. of Computer Science \& Engineering, Mississippi State University,
Starkville, MS, USA\\
\{{dp1657\IEEEauthorrefmark{1},
sm3843\IEEEauthorrefmark{2},
sn922\IEEEauthorrefmark{3}\}@msstate.edu},
mittal@cse.msstate.edu\IEEEauthorrefmark{4}}
\IEEEauthorblockA{Argonne National Laboratory,
Ankeny, IA, USA\\
{bblakely@anl.gov\IEEEauthorrefmark{5}}}}



\maketitle

\begin{abstract}
Cyberattacks are becoming increasingly difficult to detect and prevent due to their sophistication. In response, Autonomous Intelligent Cyber-defense Agents (AICAs) are emerging as crucial solutions. One prominent AICA agent is the Intrusion Response System (IRS), which is critical for mitigating threats after detection. IRS uses several Tactics, Techniques, and Procedures (TTPs) to mitigate attacks and restore the infrastructure to normal operations. Continuous monitoring of the enterprise infrastructure is an essential TTP the IRS uses. However, each system serves different purposes to meet operational needs. Integrating these disparate sources for continuous monitoring increases pre-processing complexity and limits automation, eventually prolonging critical response time for attackers to exploit. We propose a unified IRS Knowledge Graph ontology (IRSKG) that streamlines the onboarding of new enterprise systems as a source for the AICAs. Our ontology can capture system monitoring logs and supplemental data, such as a rules repository containing the administrator-defined policies to dictate the IRS responses. Besides, our ontology permits us to incorporate dynamic changes to adapt to the evolving cyber-threat landscape. This robust yet concise design allows machine learning models to train effectively and recover a compromised system to its desired state autonomously with explainability.   
\end{abstract}

\begin{IEEEkeywords}
Cybersecurity, Intrusion Response System (IRS), Knowledge Graph, Ontology, Artificial Intelligence (AI)
\end{IEEEkeywords}

\maketitle

\section {Introduction}


Most of today's automated cyber defense tools are passive watchers and do little to plan and execute responses to attacks, as well as recovery activities \cite{kottautonomous}. Response and recovery are the two core components of cyber resilience and are left for human cyber analysts, incident responders, and system administrators. Given the escalating threats, \textit{Autonomous Intelligent Cyber-defense Agents (AICAs)} have emerged as a security mechanism that offers adaptive and real-time protection against evolving digital threats. AICAs leverage AI and Machine Learning (ML) techniques to independently monitor network traffic, identify anomalies, and respond to potential security breaches without continuous human intervention. Specifically, AICA, with its Intrusion Detection System (IDS) and Intrusion Response System (IRS) components, is designed to automatically identify and initiate the most effective response to an ongoing attack \cite{ragsdale2000adaptation}. IDS identifies potential security breaches or attacks by monitoring network traffic and system activities. The IRS then dynamically adjusts its defense strategies based on the identified threat's nature, effectively diluting the impact before it can cause significant damage while restoring the system to its desired state.  

To effectively respond to an active attack, the IRS requires data from various sources, such as IDS and enterprise sensors, to identify suitable Rules of Engagement (RoEs) and determine intended behavior. This multi-source data is then used with RoEs to further train the AI and ML models for predictions, enhancing the dynamic and real-time fortifications against these attacks. As a result, the IRS must process system logs, RoEs, and AI/ML model input data for learning and prediction. However, enterprise systems possess individual schemas, leading to complex knowledge propagation and increased inference time. This multi-faceted data ingestion creates several problems (listed below) that significantly impede cyber defense operations. 

\begin{itemize}

    \item Firstly, AICAs need to interact and share information seamlessly. Diverse operating schema between multiple systems often results in data misinterpretation and complicates IRS AI/ML training and predictive modeling. Without proper training, ML systems cannot differentiate between benign and legitimate threats, increasing false positives or unnecessary alerts and responses.

    \item Secondly, cyber defense is a collaborative effort where organizations, governmental bodies, and security agencies collaborate together to address threats across organizational boundaries. Having different communication schemas to share threat intelligence and response strategies prolongs information sharing. 

    \item Lastly, the cyber threat landscape is dynamic, meaning attack vectors, tools, and defense mechanisms evolve, and the IRS must frequently adapt to new threat patterns. Without a standardized new information ingestion channel, agility is compromised, and various regulatory compliance audits for detecting, responding, and reporting cyber incidents are prevented. 
    
\end{itemize}

\begin{figure}[ht]
    \centering
    \includegraphics[scale=0.35]{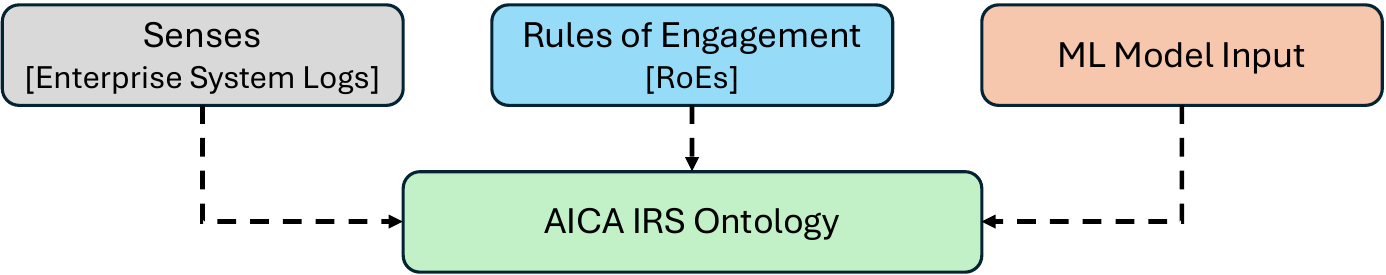}
    \caption{AICA Intrusion Response System Knowledge Graph (IRSKG) ontology to store different senses: enterprise system logs, Rules of Engagements (RoEs), and AI/ML model input.}
    \label{fig:aica-irs-ontology}
\end{figure}

Unfortunately, due to these limitations, Intrusion Response Systems (IRS) have not been able to keep up with the increasing threats \cite{cardellini2022irs}. In order to tackle this ongoing issue, we have developed a knowledge graph ontology that can encompass senses and strategies from various sources. By representing knowledge from diverse sources in a unified manner, we can enhance the AICA's efficiency in prompt response and recovery. The same ontology can also capture the RoEs and AI/ML model input data, as shown in Fig. \ref{fig:aica-irs-ontology}. Therefore, this work introduces a revolutionary AICAs IRS Knowledge Graph (IRSKG) ontology that allows streamlined knowledge ingestion, sharing, and adaptation. We demonstrate our ontology implementation utilizing a case study (see Section \ref{case-studies} for details): a network infrastructure management enterprise system. To validate our approach, we demonstrated improved IRS representation techniques for AI/ML models. Due to the graphical structure, we demonstrate a Graph Neural Network (GNN) representation for defensive cyber operations~\cite{mitra2024use} using the IRS. It is important to note that our generalized ontology is designed to accommodate any techniques, AI/ML models of choice, and enterprise systems. As per our knowledge, this research is the first attempt to develop a unified knowledge graph ontology for IRS systems. 

The rest of the paper is organized as follows - Section \ref{graphs-intro} presents background information and related work pertinent to graph notation. In addition, we introduce a prototype in Section \ref{aica-architure} that we use to demonstrate our ontology. Next, we define the ontology, AICA-IRS-KG, in Section \ref{aica-irs-kg-description}. Following the ontology definition, we present the Case Study demonstrating it by choosing the prototype in Section \ref{case-studies}. Finally, we conclude the paper in Section \ref{conclusion}.

\section {Background}
\label{graphs-intro}

In this section, we cover the preliminaries of our ontology. We cover popular graph semantics to build an ontology for the AICA IRS system. We extensively evaluate two popular graph model techniques, namely \textit{Resource Description Framework (RDF)} and \textit{Property Graph (PG)} in Section \ref{pg}. Then, we briefly describe AICA, IRS, IRS rules, IRS computation model, and AICA prototype in Section \ref{aica-architure}.\\

\subsection{Knowledge Representation, Knowledge Graph Ontologies, and Property Graphs (PG)}
\label{pg}
\textit{Knowledge Representation} (KR) organizes information in a way the computer software can understand and use to solve specific tasks. KR captures real-world knowledge so that software can process and reason. \textit{Ontologies} are specialized approaches of KR and act as blueprints to represent knowledge in specific domains. They define the entities and their relationships in a particular domain relevant only to that domain. \textit{Property Graph (PG)} \cite{tian2023world} schema handles complex and evolving relationships defined by the ontologies. We found a few ontologies on security alert systems \cite{syed2020cybersecurity}, smart city security \cite{qamar2020cyber}, information system risk management \cite{arogundade2020ontology}, cyber threat intelligence~\cite{mitra2021combating}, etc. However, to our knowledge, no specific ontology exists for cyber security response systems, such as an IRS, which represents enterprise system logs, rules, and computation models. Thus, we propose a knowledge graph called IRS Knowledge Graph (IRSKG). The PG schema specification has nodes and edges as the fundamental building blocks. These graph nodes and edges help represent entities and their relationships. 
The PG schema, also known as Labeled Property Graph (LPG) \cite{proplabelgraph}, has the following elements: \textit{`vertices'}, \textit {`edges'}, a collection of \textit {`properties'}, and \textit{`labels'}. The vertices and the edges can have only one label, while they can have multiple properties represented as key-value pairs \cite{propgraphdef}. A formal PG specification can be found at \cite {angles2018property}, which factors in a few more parameters than the four properties mentioned here. The specification uses a circle to represent a \textit{`vertex'} having a \textit{`label'} which identifies it. An arrow represents an edge between two vertices with an identity represented by a \textit{`label'}. A rectangle represents multiple key-value pairs for vertices and edges.

\begin{figure}[ht]
    \centering
    \includegraphics[scale=0.34]{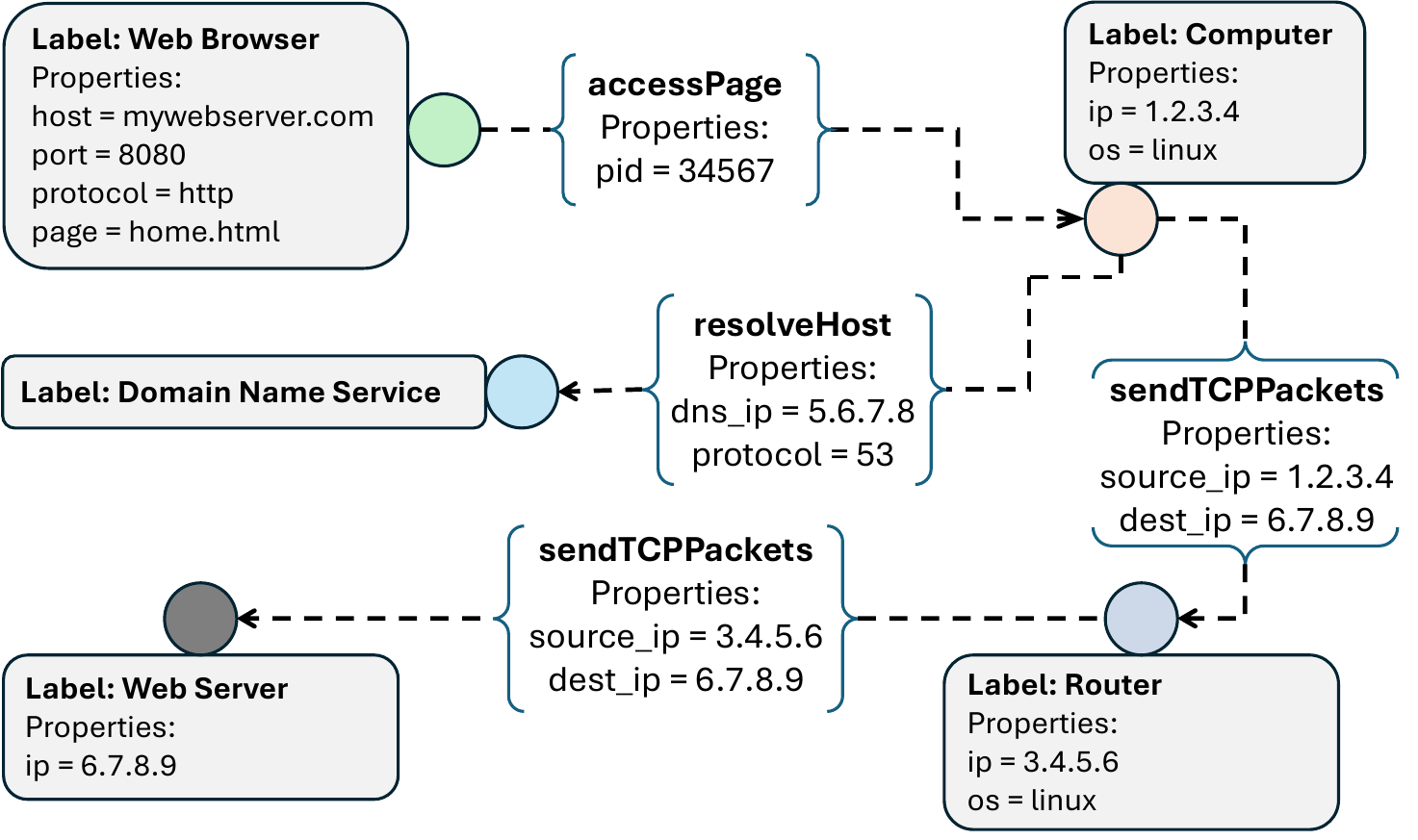}
    \caption{Graph model built using Property Graph (PG), also called Labeled Property Graph (LPG). A partial yet simple illustration of a TCP packet flow between a Web browser and an Intranet-hosted Web Server.}
    \label{fig:pg-example}
\end{figure}

We demonstrate the PG specification by showing a TCP packet flow between a Web browser and an intranet-hosted web server. Figure \ref{fig:pg-example} captures a partial network flow of the TCP packets from the Web browser to the Web server. The Graph model depicts the TCP packet flow starting from a \texttt{Web Browser} that attempts to get \texttt{home.html} hosted in \texttt{mywebserver.com} using \texttt{HTTP} protocol on port \texttt{8080}. The \texttt{Computer}, where the \texttt{Web Browser} is running, then contacts a \texttt{Domain Name Service}, to resolve the \texttt{mywebserver.com} to a destination address. The \texttt{Computer} then forwards the request to a network \texttt{Router} with the destination address, which eventually connects to the \texttt{Web Server}. An example of two vertices connected via an edge: the vertex \texttt{Web Browser} (properties: \texttt{host = mywebserver.com, port = 8080), protocol = http, page = home.html} has an edge \\ \texttt{accessPage} (properties: \texttt{pid = 34567}) to another vertex \texttt{Computer} (properties: \texttt{ip = 1.2.3.4, os = linux}).

Another specification schema is the Resource Description Framework (RDF)~\cite{RDFSpec}. It draws its inspiration from organizing information on the Web. Its genesis lies in capturing the relationships among different web pages on the Internet. Its eloquence lies in the fact that the model accommodates changes relatively quickly when a new relationship emerges without modifying a substantial portion of the graph. The graph, the linking structure, forms a labeled structure linking two resources. The first specification, RDF 1.0, was published in 1994 \cite{shadbolt2006semantic}, and soon after, Open Web Ontology (OWL) and Simple Knowledge Organization System (SKOS) specifications were built on top of RDF. We chose PG as the foundational schema for our Intrusion Response Knowledge Graph ontology over RDF for reasons that we discuss in Section \ref{aica-irs-kg-description}.

\subsection{Intrusion Response System (IRS)}
\label{aica-architure}

\begin{figure}[ht]
    \centering
    \includegraphics[scale=0.6]{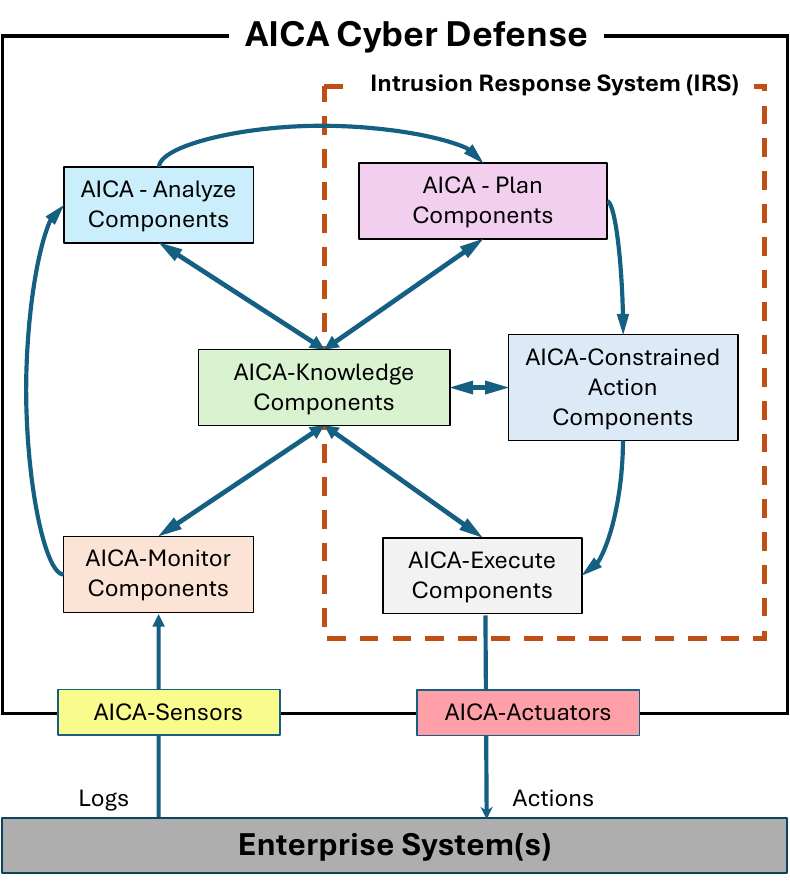}
    \caption{AICA Prototype - Self Adaptive-Autonomic Computing System based MAPE-K~\cite{1160055}(SA-ACS) framework implementation. The IRS components are responsible for recovering the enterprise system(s) to its desired state in the event of a security breach. The prototype interacts with enterprise systems via the percepts and actuators. The former gathers the logs, while the latter fixes the breached enterprise system(s). The IRS-Plan component uses the logs and the rules to create a computation model. The IRS-Constrained Action component determines the final breach mitigation action(s) following RoE.}
    \label{fig:AICA-prototype}
\end{figure}

AICAs defend enterprise systems from cyber security breaches~\cite{kott2023autonomous}. The AICAs are a group of security software that collectively continuously monitors systems and detects and remedies security breaches. One such instance is the Intrusion Response System (IRS). It restores systems to their desired state during a cyberattack, as shown in Figure \ref{fig:AICA-prototype}.

The IRS is an AICA component responsible for thwarting security breaches. It automates the procedure of responding to cybersecurity breaches to save time and reduce damage. IRS aims to prevent an attack and restore breached enterprise systems to their desired state by following certain enterprise administrators' defined IRS governing rules. The primary two IRS \textit{Plan} and \textit{Execute} components collectively fulfill IRS objectives. They use rules as a guiding principle to generate actions to thwart security breaches. Moreover, the IRS Plan component uses these rules and the system logs to train an AI/ML model that generates a response to a security breach.

To meet its goal, an IRS computes a wide \textit{variety of potential responses} including taking actions to prevent the attack from completing, restoring the system to comply with the organizational security policy, containing or confining an attack, attack eradication, deploying forensics measures to enable future attack analysis, counterattack, etc. IRS depends on governing rules defined by administrators to compute responses. There are two primary category rules governing an IRS. The \textit{first category} tells the IRS when to trigger such actions. The IRS Plan and Execute components use these rules. However, such a system must have defined rules to constrain its actions. They are the \textit{second rule category}, called Rules of Engagement (ROEs). Systems must determine which actions they can take in a fully automated manner (and when), which actions require confirmation from a human operator, and which actions must never be executed. The IRS-constrained action components use RoEs. We define the rules, their semantics, and templates, with an illustration in Section \ref{rules-data-model}.

The IRS Plan component uses different techniques, such as game theory, machine learning, etc., to create computational models. It utilizes the rules and the enterprise system information logs to build the models. The component employs these pre-trained models to generate the best response to thwart security breaches and restore the enterprise system to its desired state defined by the administrators.

We use an \textit{AICA prototype}, as shown in Figure \ref{fig:AICA-prototype}, to demonstrate our ontology implementation in the Case Study Section \ref{case-studies}. Next, we describe the prototype data flow. First, the \textit{AICA-Monitoring} managing system components continuously monitor the enterprise systems, thereby collecting the data from AICA sensors and storing data in the \textit{AICA-Knowledge} components. Second, the \textit{AICA-Analyze} components continuously investigate the stored data to detect threats. Third, the \textit{AICA-Plan} component creates a plan for each threat, to thwart cyberattacks. Fourth, the \textit{AICA-Constrained Action} component receives a response from the Plan component, checks if the response is permitted as per the RoEs, creates a final response and passes it forward. Fifth, the \textit{AICA-Execute} component executes the response on the enterprise system(s) through AICA actuators to restore the system to its desired state. 

Our IRS Knowledge Graph (IRSKG) represents enterprise system logs, rules, and computation models. In the next section, we define it, followed by the materialization of it using a Case Study \ref{case-studies} using AICA Prototype.

\section{Intrusion Response System Knowledge Graph (IRSKG) Ontology Schema}
\label{aica-irs-kg-description}

\begin{figure*}[ht]
    \centering
    \includegraphics[scale=0.58]{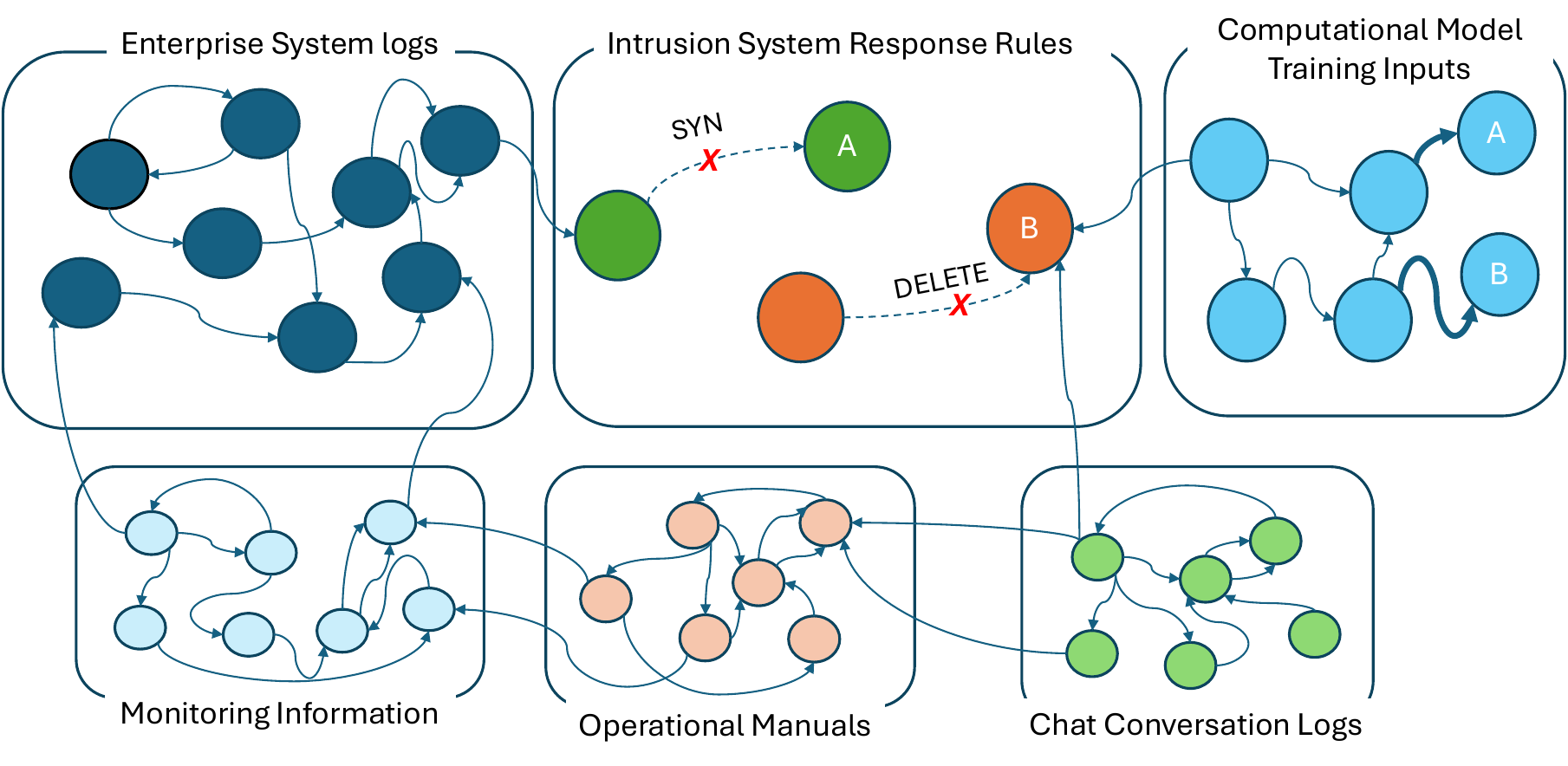}
    \caption{Illustration of IRSKG: A Graph-based model that represents enterprise system information such as System logs, System monitoring logs, Chat conversation logs, IRS rules, and input data for the computational model training.}
    \label{fig:AICA-IRS-KG-Overview}
\end{figure*}

In this section, we introduce a unified ontology to capture the semantic relationships among different components of an Intrusion Response System (IRS). In addition, we create a schema called IRSKG to represent the ontology in a data structure. Our schema, shown in figure \ref{fig:AICA-IRS-KG-Overview}, has been designed to represent disparate enterprise system information such as system logs, system monitoring information, chat conversation logs, intrusion response rules, and response computation model input data. To the best of our knowledge, there is no publicly available ontology that helps represent the above mentioned data for Intrusion Response Systems (IRSs). To build our IRKG ontology we utilize the \emph{PG} specification (See Section \ref{pg}) over \emph{RDF} because of the following primary reasons:
\begin{itemize}
	\item PG is widely adopted in cyber security domain \cite{angles2023pg}, and is more suitable for dynamic datasets \cite{rdfvspg1}.
	\item PG uses the flexible and extensible JSON format, unlike RDF, which uses XML.
 	\item Information retrieval semantic standard is available for PG \cite{deutsch2022graph}.
\end{itemize}
In our IRSKG case studies (Section \ref{case-studies}), we utilize Neo4J \cite{neo4j}, a PG, 
to demonstrate our ontology implementation. 
Next, we explain the IRSKG ontology schema in detail. For simplicity we only discuss three enterprise system information schema. The following three subsections describe the schema for the enterprise systems logs, IRS rules, and the computation model inputs. 
Table \ref{tbl:notation-table} summarizes the used set of notations.

\begin{table}
    \footnotesize
    \renewcommand{\arraystretch}{1.30}%
    \caption{IRSKG schema notations represent disparate enterprise system information such as system logs, system monitoring information, chat conversation logs, intrusion response rules, and response computation model input data.}
    \begin{tabularx}{0.48\textwidth} { 
       >{\centering\arraybackslash}p{0.10\textwidth}|X
       >{\raggedright\arraybackslash}X }
        \hline
            \rowcolor{lightgray} 
            \textbf{Symbol} &  \textbf{Description} \\
        \hline
            $\G$ & The entire Graph database \\
            $\V$ & The set of all vertices in the database. \\ 
            $\E$ & The set of all edges in the database \\ 
            $\V_i$ & The $ith$ vertex. \\ 
            $\E_{i,j}$ & The edge between the $\V_i$ and $\V_j$\\ 
            $\lb(\V_i)$ & The label of $\V_i$. \\ 
            $\pr(\V_i)$ & The property key-value dictionary of $\V_i$\\ 
            $\lb(\E_{i,j})$ & The label of $\E_{i,j}$ edge. \\ 
            $\pr(\E_{i,j})$ & The property key-value dictionary of $\E_{i,j}$ edge\\  
            $\RoE$ & Set of all Rules of Engagement \\  
            $\RoE_i$ & $ith$ Rule of Engagement\\
            $\V_{a|b}(\R_i)$ & Vertex $a$ or $b$ of rule $\RoE_i$\\  
            $\lb(\V_a(\R_i))$ & Label of the Vertex $a$ of $\R_i$\\  
            $\pr(\V_a(\R_i))$ & Property of the Vertex $a$ of $\R_i$\\  
            $\E({\RoE_i})$ & Edge between $\V_a$ and $\V_b$ of $\RoE_i$\\  
            $\lb(\E({\RoE_i}))$ & Label of edge between $\V_a$ and $\V_b$ of $\RoE_i$\\  
            $\pr(\E(\RoE_i))$ & Property of edge between $\V_a$ and $\V_b$ of $\RoE_i$\\  
            $\RoE^{t} \ | \ \RoE^i \in \RoE^{ti}$ & A meta-template that different enterprise systems follow and ultimately all $\RoE_{i}$ are compliant to.\\
            $\RoE^{tk}$ & A template for a specific enterprise system $k$ (e.g. A web enterprise system). \\ 
        \hline  
    \end{tabularx}
    \label{tbl:notation-table}
\end{table}

\subsection{Enterprise System Log Schema}  
\label{aica-data-model}
In this section, we create a schema that captures the logs of heterogeneous enterprise systems. Typically, the logs are in single-line text entries and are better suited for anything but unstructured textual repository format. We choose a graphical representation model to relate information visually. The IRSKG schema uses the graph notations and represents a \textit{`Graph'} ($\G$) as a set of \textit{`vertices'} ($\V$) and \textit{`edges'} ($\E$) as shown in the Equation \ref{eq:graph-ve}. We use the PG model technique (see section \ref{pg}) and thus each vertex and edges have one or more \textit{`properties'} along with a \textit{`label'} that identifies the vertex or edge as shown in the Equations \ref{eq:graph-v} and \ref{eq:graph-e} respectively.

\begin{subequations}\label{property-graph}
    \begin{equation} \label{eq:graph-ve}
             \G = \ (\V, \ \E) \ | \ \E_{i,j} \in \E,  \ \V_i \in \V 
    \end{equation} 
    \begin{equation} \label{eq:graph-v}
        \begin{array} {ll}
             \V_i = \{\lb(\V_i), \ \pr(V_i) \} \\
        \end{array}    
    \end{equation}     
    \begin{equation} \label{eq:graph-e}
        \begin{array} {ll}
             \E_{i,j} = \{ \lb(\E_{i,j}), \pr(\E_{i,j})\}) \\
        \end{array}    
    \end{equation}     
\end{subequations}\\

We demonstrate an implementation in Section \ref{case-study-senses-graph-model}.

\subsection{Intrusion Response System Rules Schema} \label{rules-data-model}

We create a schema to represent all the IRS governing rules. The rules are the instructions that tell the IRS how to thwart a security breach. The rules comprise two primary categories: the first category is conditions that trigger the IRS to take action on the breach, and the second is constraints that overwrite the IRS actions deemed unsafe to the system. Usually, these rules follow different schemas and are stored in various file formats (JSON, XML, YARA specification, etc.) IRSKG enables a unified graph schema to represent the rules that show the relationship among the components of the various rules. Next, we describe the IRS governing rules graph notations, their semantics, templates, and constraints, followed by their illustration.\\

\subsubsection{Rules Graph Notation}
$RoEs$ influence the AICA IRS Plan Components ML models that predict an action or a set of actions to thwart the security breach and restore the enterprise systems to their predefined desired state. We denote RoE as a set of rules $\RoE_i$, as shown in the Equation \ref{eq:rule-set}. $\RoE_i$ has two vertices and an edge connecting those vertices as shown in Equation \ref{eq:rule-graph-ve}. We represent the vertices as $\V_{a|b}(\RoE_i)$, with a label, $\lb(V_a(\RoE_i))$,  and a property set, $\pr(\V_a(\RoE_i))$, as shown in Equation \ref{eq:rule-graph-v}. In addition, we express the edge between $\V_{a}(\RoE_i)$ and $\V_{b}(\RoE_i)$ as $\E(\RoE_i)$, with a label $\lb(\E(\RoE_i))$ and a property set $\pr(\E(\RoE_i))$ as shown in Equation \ref{eq:rule-graph-e}. The property set is a \{key, value\} pair and can have any arbitrary number of such pairs. However, the organization admin constrains the property set by defining a template ($\RoE^t$) that all $\RoE_i$ must comply with. The logic can be related to the inheritance concept in object-oriented programming for improved understanding.\\

\begin{subequations}\label{rulesengine-graph}
    \begin{equation} \label{eq:rule-set}
        \begin{array} {ll}
             \RoE = \{{\RoE_i \ \forall \ 0 < i < n\}} 
        \end{array}  
    \end{equation}         
    \begin{equation} \label{eq:rule-graph-ve}
        \begin{array} {ll}
             \RoE_i = \{{\V_a(\RoE_i), \E(\RoE_i), \V_b(\RoE_i)\}} 
        \end{array}    
    \end{equation} 
    \begin{equation} \label{eq:rule-graph-v}
        \begin{array} {ll}
             \V_a(\RoE_i) = \{ \lb(\V_a(\RoE_i)), \pr(\V_a(\RoE_i)) \} 
        \end{array}    
    \end{equation}     
    \begin{equation} \label{eq:rule-graph-e}
        \begin{array} {ll}
             \E(\RoE_i) = \{ \lb(\E(\RoE_i)), \ \pr(\E(\RoE_i))) 
        \end{array}    
    \end{equation}     
\end{subequations}\\

\subsubsection{Rule Symantic}
\label{rule-symantic}
The $\RoE_i$ captures \textit{`who can do what in which resource'}. The \textit{`who'} and \textit{`which'} are vertices and capture the source and destination computers. The \textit{`who'} and \textit{`which'} are vertices, $\V_{a}(\R_i)$ and $\V_{b}(\R_i)$ respectively. We capture the computer \textit{`IP'} as the vertex label, namely, $\lb(\V_a(\R_i))$ and $\lb(\V_b(\R_i))$. However, our schema is flexible; thus, it can contain additional properties such as computer name, computer location, asset name, asset tag, etc. We capture these in $\pr(\V_a(\R_i))$ and $\pr(\V_b(\R_i))$. The \textit{`what'} is a verb that the \textit{`who'} wants to carry on \textit{`which} resources. We represent the \textit{`what'} as en edge, $\E({\RoE_i})$ between $\V_a(\R_i)$ and $\V_b(\R_i)$. The $\lb(\E({\RoE_i}))$, the label captures the verb. The $\E({\RoE_i})$ relationship also captures the \textit{`constraint'} that  $\E({\RoE_i})$ should follow as laid out by the organization administrators.\\

\subsubsection{Rule constraint}
We further define the \textit{`rule constraints'} that organization administrators define to \textit{`allow'} or \textit{`deny'} certain graph relationships. As an illustration, a certain path from vertex $\V_{a}(\R_i)$ to vertex $\V_{b}(\R_i)$ with an edge $\E({\RoE_i})$ is not allowed. We capture this value as a property in the $\E({\RoE_i})$ with a key \textit{`constraint'}. We use this while training the GNN to give a special weight to the relationships of the graph. We aim to have the GNN predict the appropriate action, either deny or allow at model inference time. In addition, our notations are flexible to accommodate additional constraints as needed by the enterprise systems.\\

\subsubsection{Rules Template}
We use \textit{`meta-template'} and specific \textit{`templates'} for each enterprise system that each rule should comply with. The \textit{`meta-template'} governs all enterprise system templates and hence all rules. Each rule must follow an enterprise system template that also adheres to the meta templates and introduces further semantics specific to that enterprise system. We define $\RoE^t$ as the meta template that adheres to the $\RoE_i$ as we explain in Equation \ref{eq:rule-graph-ve}. We define enterprise system templates as  $\RoE^{tk}$. The organization administrator specifies the necessary rule semantics in the enterprise system template. Each enterprise system has precisely one enterprise system template. All instance of enterprise system uses the same system template. Each enterprise system template, $\RoE^{tk}$, defines a rule set, $\RoE^{tk}(j)$. Two enterprise systems differ at the least by 'what' (verb) allowed on them. We further explain the template mechanism with concrete examples later in the Case Study Section \ref{case-studies}.\newline 

\subsubsection{Rules illustration}
We illustrate a RoE that uses our IRSKG notation in this section. One rule in $\RoE^{t1}$ could represent a consolidated set of rules $\RoE^{t1}_i$. As an illustration, an administrator might want to deny remote connectivity from any source IP to a critical network asset such as a physical router in a Network Infrastructure management enterprise system. We represent such a rule with the source IP of \textit{`any'} as the $\lb(\V_a(\RoE_1)$ of the vertex $\V_a(\RoE_1)$, the destination IP consists of the router IP as $\lb(\V_b(\RoE_1)$ of the vertex $\V_b(\RoE_1)$, and the connectivity \textit {`SYN'} as $\lb(\E(\RoE_1))$ with a property \textit{`constraint'} having value \textit {`deny'} as shown in the Equation \ref{eq:network-router-example}. The equation adheres to the model we describe in Section \ref{eq:rule-set}. Moreover, it also adheres to the semantics we define in Section \ref{rule-symantic} where \textit{`who'} maps to $\lb(\V_a(\RoE_1))$, \textit{`which'} maps to $\lb(\V_b(\RoE_1))$, \textit{`what'} maps to the verb $\lb(\E(\RoE_1))$, and the property \textit{`constraint'} with value \textit {`deny'} is represented in $\pr(\E(\RoE_1))$ as $\{``constraint": \ ``deny"\}$.

\begin{subequations}\label{eq:network-router-example}
    \begin{equation} \label{eq:router-roe1}
        \begin{array} {ll}
             \RoE_1 = \{{\V_a(\RoE_1), \  \E(\RoE_1), \ \V_b(\RoE_1)\}} 
        \end{array}  
    \end{equation}  
    \newline
    \begin{equation} \label{eq:router-roe1-v1}
        \begin{array} {ll}
             \V_a(\RoE_1) = \{ \lb(\V_a(\RoE_1)), \pr(\V_a(\RoE_1))\} \\ | \  \lb(\V_a(\RoE_1)) \ \textrm{\textit{``any"}}, \ \pr(\V_a(\RoE_1)) \textrm{ = \{{...}\} }
        \end{array}    
    \end{equation} 
    \newline
    \begin{equation} \label{eq:router-roe1-v2}
        \begin{array} {ll}
             \V_b(\RoE_1) = \{ \lb(\V_b(\RoE_1)), \pr(\V_b(\RoE_1))\} \\ 
             | \lb(V_b(\RoE_1)) = ``1.2.3.4", \pr(\V_b(\RoE_1)) \textrm{ = \{{...}\} }
             
        \end{array}    
    \end{equation}   
    \newline
    \begin{equation} \label{eq:router-roe1-e}
        \begin{array} {ll}
             \E(\RoE_1) = \{ \lb(\E(\RoE_1)),\pr(\E(\RoE_1))\} \ | \ \textrm{$\lb(\E(\RoE_1))$ =\textit{`SYN'},} \\
             \pr(\E(\RoE_1)) = \textrm{\{\textit{`constraint'}: \textit{`deny'} \}} 
        \end{array}    
    \end{equation}     
\end{subequations}

In Section \ref{use-case-rules-graph-model} we demonstrate schema implementation.\\

\subsection{Response Computation Model Input Data Schema} \label{data-transformation}
In addition to the logs and the rules, IRSKG creates a schema for the input data that the IRS uses to prepare a response computational model. IRS uses the model to create action to thwart security breaches. We use a graph neural network (GNN) as a machine learning model to demonstrate the transformation concept. One can change to any other format suitable for a different machine learning model. Next, we elaborate on IRSKG notations on this ML model input data in this section.

We aggregate the cumulative outbound and inbound connections from and to from each vertex, $\V_i$, and represent them in a property in $\pr(\V_i)$, \textit{`count'} as shown in the Equation \ref{eq:gnn-sensor-graph-transform-vertex}. In addition, we also represent the total connections between two vertices, $\V_i$, and $\V_j$ in their edge, $\E_{i,j}$, in a property in the dictionary, $\pr(\E_{i,j})$, called \textit {`count'} shown in Equation \ref {eq:gnn-sensor-graph-transform-edge}.

\begin{subequations}\label{eq:gnn-sensor-graph-transform}
    \begin{equation} \label{eq:gnn-sensor-graph-transform-vertex}
        \begin{array} {ll}
             \pr_{count}(\V_i)  = deg(\V_i) \ | \ deg \textrm{\ = degree of vertex $\V_i$}\\
        \end{array}  
    \end{equation}
    \newline
    \begin{equation} \label{eq:gnn-sensor-graph-transform-edge}
        \begin{array} {ll}
             \pr_{count}(\E_{i,j}) = \pr_{count}(\V_i) + \pr_{count}(\V_j)
        \end{array}    
    \end{equation}     
\end{subequations}

Next, we provide data transform examples that is needed to create a GNN model, using Equation \ref{eq:gnn-sensor-graph-transform-vertex} to calculate $\pr_{count}(\V_i)$. Its value is either a cumulative count value as explained by the equation or is a hyper-parameter. For the latter, for example, we transform the constraint rule $\RoE_1$ defined in the Equation \ref{eq:network-router-example}  as follows: the rule $\RoE_1$, with a vertex label $\lb(\V_1(\RoE_1))$ of \textit{`any'} with an edge $\E(\RoE_1)$ with a property key and value pair represented in $\pr(\E(\RoE_1))$ as \{\textit{`constraint'}: \textit {`deny'}\} to a $\pr_{count}(\V_1) = -100000$. The -1000000 value, is a data model hyper-parameter set at the design time, makes the network ignore the edge between any vertex to the vertex $\V_a(\RoE_1)$ as shown in the Equation \ref{eq:gnn-sensor-graph-transform-edge-constraint}. The value has to be a sufficiently large negative value. The absolute value depends on the enterprise systems transformed graph  $maximum | \ \pr_{count}(\V_i) \ |$. We use GNN, so we must adopt this mechanism to define large negative values. However, one can change the mechanism to a different one based on the technique suitable for their IRS of choice.

\begin{equation} \label{eq:gnn-sensor-graph-transform-edge-constraint}
    \begin{array} {ll}
         \pr_{count}(\V_i)  = -1000000 \ | \ \lb(\V_i(\RoE_1) = \textrm{\textit{`any'},}
         \\ \textrm{$\pr(\E(\RoE_1))$ = \{\textit{`constraint'}: \textit {`deny'}\}} 
    \end{array}    
\end{equation}     

We demonstrate an implementation in Section \ref{case-study-transformation-graph-model}.\\

\section{Case Study of the Knowledge Graph for Intrusion Response System (IRS)} \label{case-studies}

\begin{figure}[ht]
    \centering
    \includegraphics[scale=0.34]{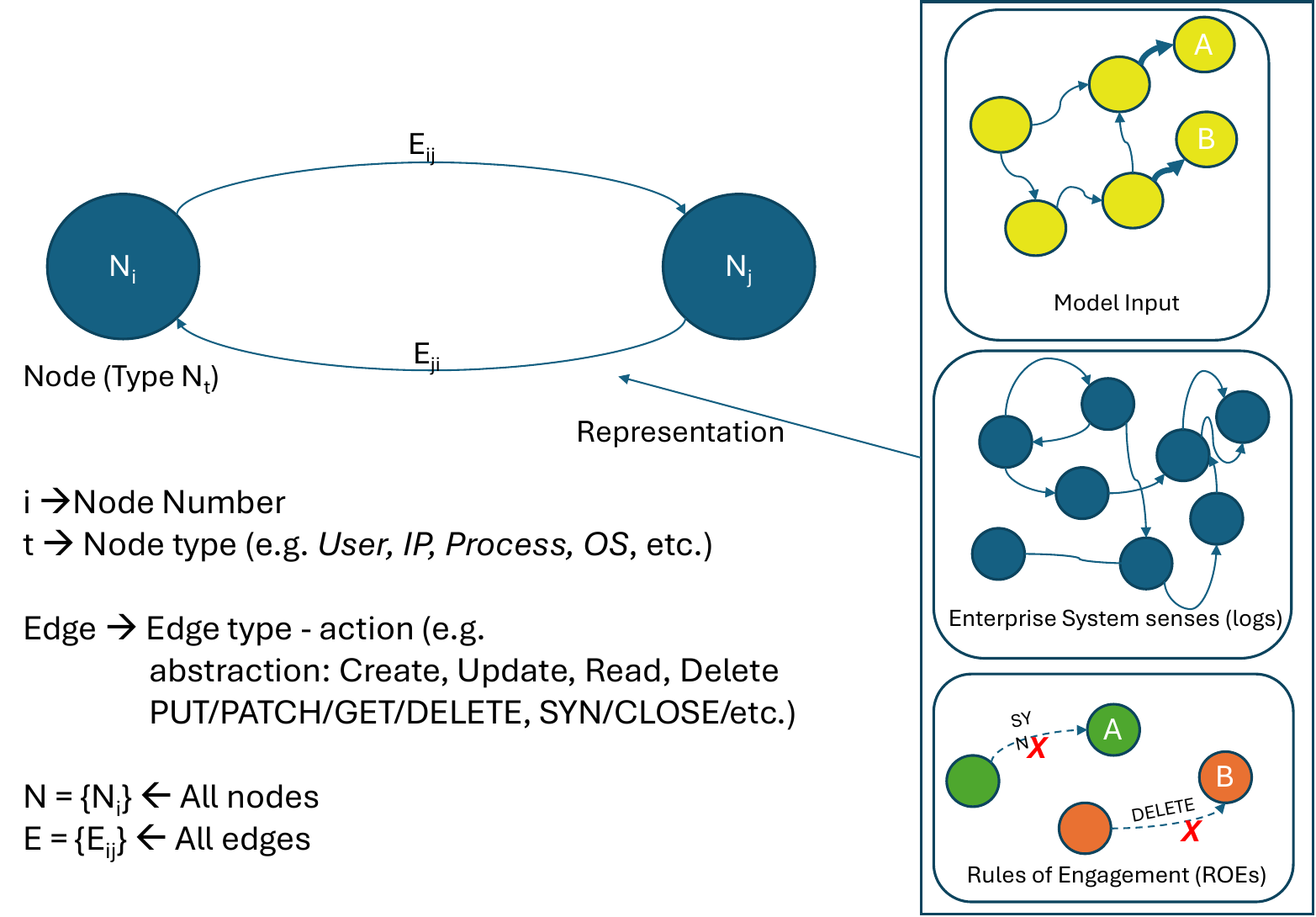}
    \caption{A generalized Network System IRSKG: represents network logs, IRS rules, and the computation model input to train a GNN. For example, for the network logs, the IRSKG \textit{Nodes}, $N_i$, represents the source and the destination IPs. \textit{Edges}, $E_{ij}$, represents the action the source wants to take on the target, e.g., \textit{SYN} represents an action when a source wants to connect with a destination.}
    \label{fig:AICA-IRS-KG-NMS}
\end{figure}

This section demonstrates an IRSKG implementation that translates the abstract schema semantics created in section \ref{aica-irs-kg-description} to software artifacts. We do so for a cyber defense case study, a Network Infrastructure Management (NIMS) enterprise system. Furthermore, we use the AICA prototype (see section \ref{aica-architure}) as the software stack to demonstrate the IRSKG representation. Moreover, we demonstrate the IRS rules, specifically the $RoE$ \textit{`constraints'} rules, as explained in Section \ref{rules-data-model}. In addition, we chose a GNN that the IRS uses as a response computation model (see section \ref{data-transformation}). Fig. \ref{fig:AICA-IRS-KG-NMS}, demonstrates a generalized IRSKG of a typical network system, RoEs that govern the network path between hosts, and the model input to train the GNN. We choose Neo4J \cite{neo4j} to demonstrate a concrete IRSKG schema semantic implementation.

Next, we demonstrate the IRSKG implementation of the NIMS that uses Graylog system logs in three tasks. In each task, we represent the subject in IRSKG, followed by its concrete implementation in Neo4J. We handle Graylog in the first task. We show raw logs, followed by their IRSKG representation and Neo4J implementation. In the second task, we illustrate an IRS rule that denies modifying the router entries. Finally, we show an IRS GNN computation model input that the IRS uses to formulate a response to thwart security breaches. We demonstrate the three above tasks in detail in the following three subsections.
\newline \newline

\subsection{Enterprise System Network Logs Graph Schema} \label{case-study-senses-graph-model}
This section demonstrates the implementation of the IRSKG enterprise system log schema (see section \ref{aica-data-model}) in Neo4J. We store NIMS logs in Neo4J, which complies with the IRSKG enterprise system schema. The current implementation of the AICA prototype uses Graylog as a Security Information and Event Management (SIEM) to consolidate logs from sensors in the environment for further IRS use. We use the logs to demonstrate our IRSKG implementation. $N_i$ represents the sources and destinations of a network log entry. $E_j$ represents the action that the source wants to take on the destination. For example, $SYN$ is an action when a source sends a connect request to a destination. We represent the activity as the source node, $NEP_1$, has $E_{1,2}$ relationship of type $SYN$ with a destination $NEP_2$. 

\begin{figure}[]
    \centering
    \includegraphics[scale=0.4]{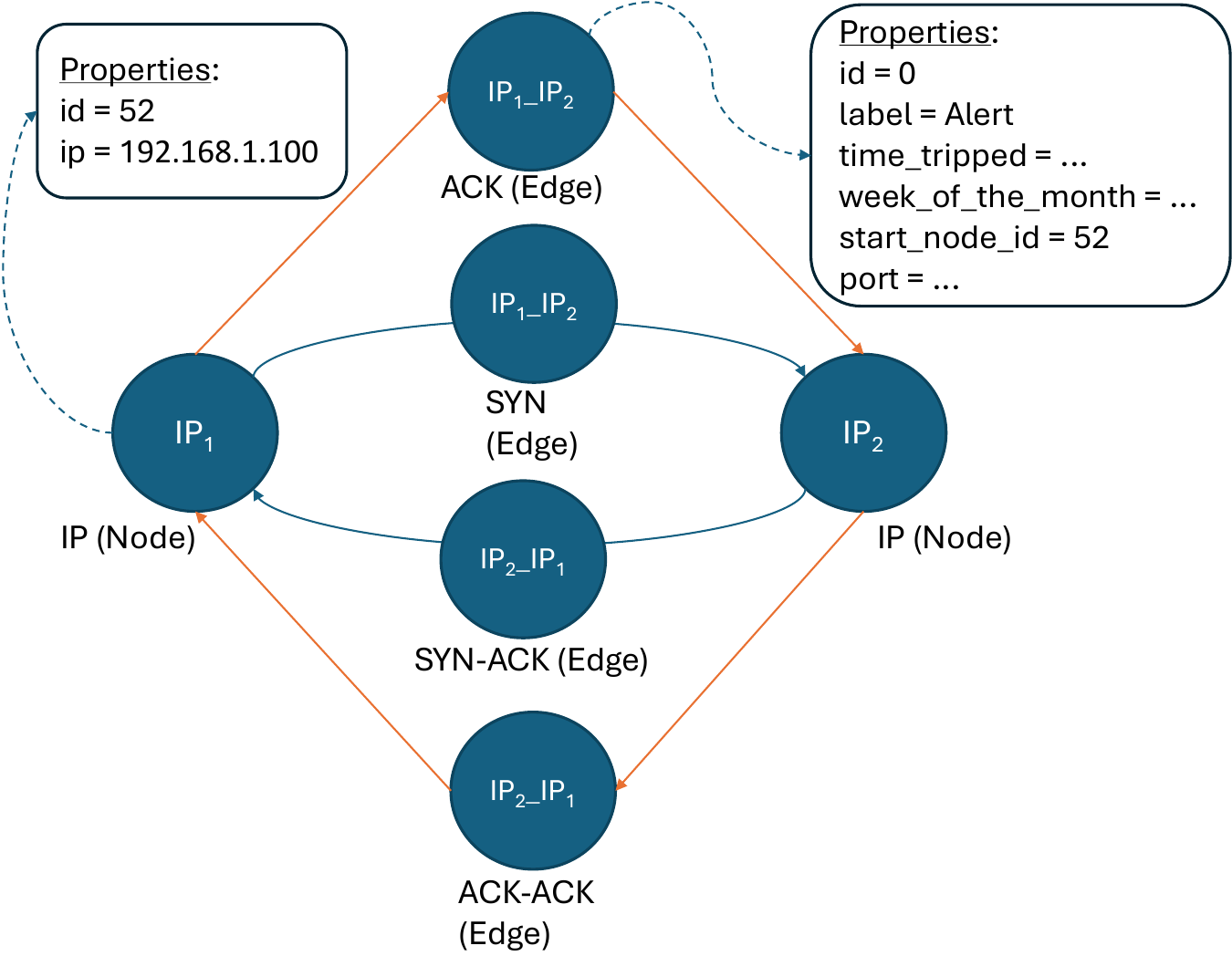}
    \caption{A Network Management Enterprise System IRSKG with sample raw Graylog entries as shown in Listing \ref{list:network-case-study-ingestion-phase-dataingestion-graylog-neo4j} that complies with IRSKG  schema as shown in Fig. \ref{fig:AICA-IRS-KG-NMS}. The figure shows that source, $IP_1$, has a property $\lb(\V_1) = ip$ with a value $\pr(V_i)= \{ ``ip": 192.168.1.100 \}$ has relationships, $SYN$ and $SYN_ACK$, with destination $IP_2$ with IP property value $192.168.1.101$.}
    \label{fig:AICA-IRS-KG-NMS-Senses}
\end{figure}

\begin{lstlisting} [caption={Sample raw Network Management System GrayLog entries. The entries show the communication between the source, \texttt{192.168.0.100}  and the destination, \texttt{192.168.0.101}. The source sends a \texttt{SYN} to the destination; it then sends back \texttt{SYN-ACK} to the source.}, label=list:network-case-study-ingestion-phase-dataingestion-raw-graylog]
[2023-10-25 11:10:45] 192.168.1.100 -> 192.168.1.101: TCP SYN 
[2023-10-25 11:10:46] 192.168.1.101 -> 192.168.1.100: TCP SYN-ACK 
[2023-10-25 11:10:47] 192.168.1.100 -> 192.168.1.101: TCP ACK
[2023-10-25 11:10:48] 192.168.1.101 -> 192.168.1.100: TCP ACK
\end{lstlisting}

We show in the Listing \ref{list:network-case-study-ingestion-phase-dataingestion-raw-graylog} a few raw Graylog entries from the Router. The Graylog listing shows that source \texttt{192.168.1.100} sends a connect, \texttt{SYN}, to the destination \texttt{192.168.1.101}. \texttt{SYN} is the first byte sent by the source. The destination responds to the source by \texttt{SYN-ACK}. Next, the source sends an \texttt{ACK} to the destination, and the latter also reciprocates with an \texttt{ACK}. We represent the log information in IRSKG as shown in Fig. \ref{fig:AICA-IRS-KG-NMS-Senses} per the schema created in Section\ref{aica-data-model}.\newline

\begin{lstlisting} [caption={IRSKG Graylog Network Management system implementation of the raw entries as shown in Listing \ref{list:network-case-study-ingestion-phase-dataingestion-raw-graylog} and Figure \ref{fig:AICA-IRS-KG-NMS-Senses}. The graph entries show two nodes with $id=0$ and $id=1$ with their corresponding ips representing the source and the destination, \texttt{192.168.1.100} and \texttt{192.168.1.101}. There is a relationship between the two of type \texttt{SYN} and with properties that store the time when source sent a \texttt{SYN} request to the destination.}, label=list:network-case-study-ingestion-phase-dataingestion-graylog-neo4j]
{"type":"node","id":"0","labels":["IP1"],"properties":{"ip":"192.168.1.100"}}
...
{"type":"node","id":"1","labels":["IP2"],"properties":{"ip":"192.168.1.101"}}
...
{"type":"relationship","id":"0","label":"SYN","start":{"id":"0","properties":{"time":"23:10:45,"time_month":10,"time_year":2023,"time_date":25},"end":{"id":"1","properties":...}}
\end{lstlisting}

\subsection{Enterprise System Network Rules Graph Schema}\label{use-case-rules-graph-model}

\begin{figure}[ht]
    \centering
    \includegraphics[scale=0.4]{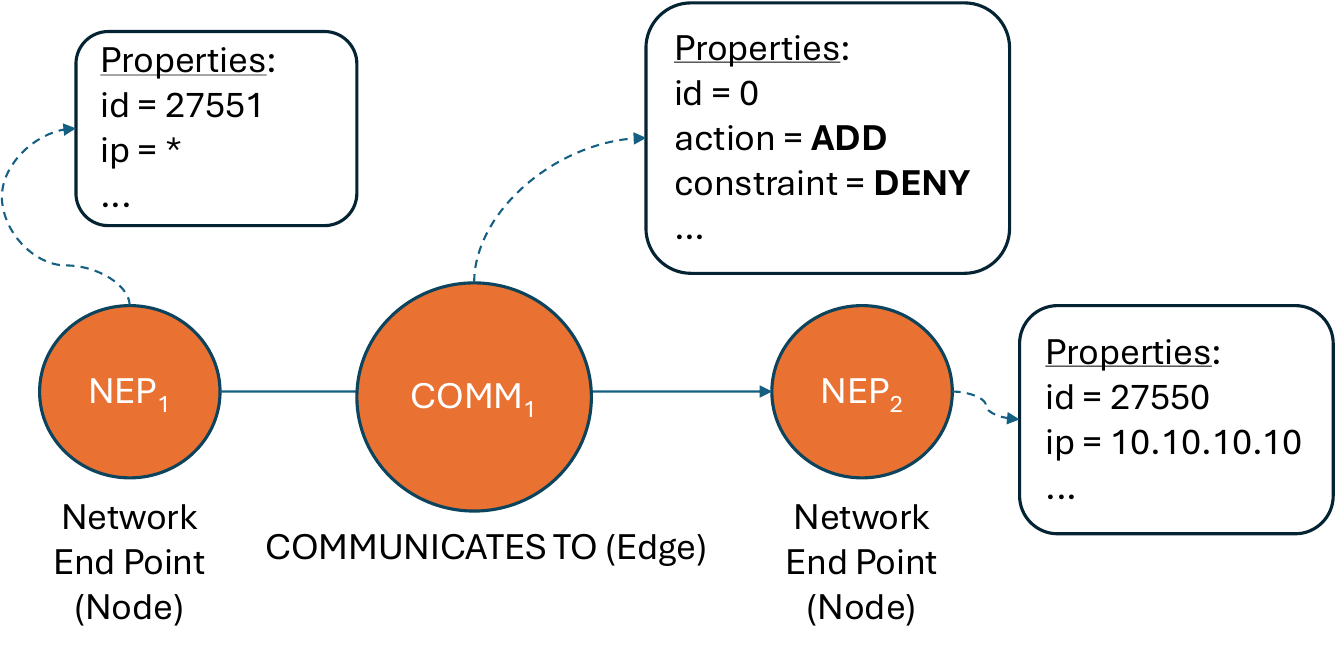}
    \caption{IRSKG for Network Management System constraint rule that denies any source, represented as $ip=*$ to modify the router represented as $ip=10.10.10.10$ with a rule to \textit{constraint} to deny (with a value \texttt{DENY}) for  an \textit{action} = \texttt{ADD} as shown in Listing \ref{list:network-case-study-ingestion-phase-dataingestion-neo4j-fw_rule} and that complies with IRSKG ontology schema as shown in Figure \ref{fig:AICA-IRS-KG-NMS}. }
    \label{fig:AICA-IRS-KG-NMS-ROE}
\end{figure}

We demonstrate the IRSKG rules schema (see Section \ref{rules-data-model}) in this section. We chose an example IRS rule, $\RoE_1$, that denies any machine (source) to modify the router with an IP 10.10.10.10. We represent the $\RoE_1$ in IRSKG as shown in Figure \ref{fig:AICA-IRS-KG-NMS-ROE} and store the rule in the Neo4J IRSKG implementation as shown in Listing \ref{list:network-case-study-ingestion-phase-dataingestion-neo4j-fw_rule}. The node type of the source and destination nodes, represented as $\lb(V_a(\RoE_1))=NEP_1$ and $\lb(V_b(\RoE_1))=NEP_2$, are \texttt{NetworkEndpoint} type. They have property name as $\textit{ip-address}$ values as $*$ and $10.10.10.10$ represented in $\pr(\V_a(\RoE_1))$ and $\pr(\V_b(\RoE_1))$ respectively. Furthermore, $NEP_1$ and $NEP_2$ has \textit{id} property, specific to Neo4J software, as $27550$ and $27551$ respectively. The edge, $\E_{1,2}$ of type \texttt{relationship}, between $NEP_1$ (\texttt{start}) and $NEP_2$ (\texttt{end}) has a label $\lb(\E(\RoE_1)) = COMMUNICATES\_TO$ with properties such as $\pr_{constraint}(\E(\RoE_1))$ as \texttt{deny}.

\begin{lstlisting} [language=YARA, mathescape=true, caption={Neo4J IRSKG IRS rule that prevents any machine from modifying the router with a \textit{constraint} \texttt{deny} for the \textit{action} \texttt{ADD} as shown in Figure \ref{fig:AICA-IRS-KG-NMS-ROE}. The source is represented as a node with \textit{id=27751} having \textit{ip\_address = *}. The router is repsenented as a node with \textit{id=27750} having \textit{ip\_address = 10.10.10.10}. The relationship is represented as an edge with an \textit{id=0} with \textit{label=COMMUNICATES\_TO}.},
label=list:network-case-study-ingestion-phase-dataingestion-neo4j-fw_rule]
{
  "type": "node",
  "id": "27551",
  "labels": [
    "NEP1"
  ],
  "properties": {
    "ip_address": "*"
  }
}
{
  "type": "node",
  "id": "27550",
  "labels": [
    "NEP2"
  ],
  "properties": {
    "ip_address": "10.10.10.10"
  }
}
{
  "type": "relationship",
  "id": "0",
  "label": "COMM1",
  "start": {
    "id": "27551",
    "labels": [
      "NEP1"
    ],
    "properties": {
      "ip_address": "*"
    },
  "end": {
    "id": "27550",
    "labels": [
      "NEP2"
    ],
    "properties": {
      "ip_address": "10.10.10.10"
    }
  }
  },
    "properties": {
      "action": "ADD",
      "constraint": "deny",      
      "id": "6ec4f95c-f4e3-4516-92c1-172cec275696"
    }
  }
}
\end{lstlisting}

\subsection{Enterprise System Network Computational Model Input Graph Schema} \label{case-study-transformation-graph-model}
Finally, we demonstrate the IRSKG response computational model input (see Section \ref{data-transformation}) implementation in this section. The IRS uses this data in the NIMS case study to train a GNN model. Next, we explain the GNN computational model input schema using the Graylogs, as shown in Fig. \ref{fig:AICA-IRS-KG-NMS-Senses} and constraint rules, as shown in Fig. \ref{fig:AICA-IRS-KG-NMS-ROE}. We illustrate how the input Gray logs, as shown in the Listing \ref{list:network-case-study-ingestion-phase-dataingestion-raw-graylog}, are transformed to the GNN input data schema as shown in the Listing \ref{list:network-case-study-ingestion-phase-datatransform-neo4j-vertex-count}. We calculate the \textit{`count'} property value, ~$\pr_{count}(\V_i)$ as $4$, following the Equation \ref{eq:gnn-sensor-graph-transform-vertex} because there are two nodes, where the ip, \textit{`192.168.1.100'} appears as either the \textit{`source'} or \textit{`destination'} address in Listing \ref{list:network-case-study-ingestion-phase-dataingestion-raw-graylog}. The source node is represented as $IP_1$ and the destination node as $IP_2$ as shown in Figure \ref{fig:AICA-IRS-KG-NMS-Senses}. Similarly, we assign the same \textit{`count'} value to the ip \textit{`192.168.1.101'} as shown in the Listing \ref{list:network-case-study-ingestion-phase-datatransform-neo4j-vertex-count}. Moreover, we calculate the property value, ~$\pr_{count}(\E_{i,j})$ as two of the edge, $\E_{1,2}$ between $IP_1$ and $IP_2$, as shown in the Listing \ref{list:network-case-study-ingestion-phase-datatransform-neo4j-edge-count} abiding to the Equation \ref{eq:gnn-sensor-graph-transform-edge}.

\begin{lstlisting} [language=YARA, mathescape=true, caption={Neo4J IRSKG computational model input implementation for vertexes $IP_1$ with $ip\_address=192.168.1.100$ and $IP_2$ with $ip\_address=192.168.1.101$: since these IPs appear twice in Graylog Listing \ref{list:network-case-study-ingestion-phase-dataingestion-raw-graylog}. Hence ~$\pr_{count}(V_i)$ as $4$ as represented in \textit{properties} \texttt{count} based on Equation \ref{eq:gnn-sensor-graph-transform-vertex}.}, label=list:network-case-study-ingestion-phase-datatransform-neo4j-vertex-count]
{
  "type": "node",
  "id": "2891",
  "labels": [
    "IP1"
  ],
  "properties": {
    "ip_address": "192.168.1.100",
    "count": 2
  }
}
{
  "type": "node",
  "id": "2892",
  "labels": [
    "IP2"
  ],
  "properties": {
    "ip_address": "192.168.1.101",
    "count": 2
  }
}
\end{lstlisting}

\begin{lstlisting} [language=YARA, mathescape=true, caption={Neo4J IRSKG computational model input implementation for the Edge, $E_{1,2}$ with two nodes - $IP_1$ with $ip\_address=192.168.1.100$ and $IP_2$ with $ip\_address=192.168.1.101$: since these IPs appear twice in Graylog Listing \ref{list:network-case-study-ingestion-phase-dataingestion-raw-graylog}. Hence ~$\pr_{count}(\E_{1,2})$ as $4$ as represented in \textit{properties} \texttt{count} based on Equation \ref{eq:gnn-sensor-graph-transform-edge}.\\}, label=list:network-case-study-ingestion-phase-datatransform-neo4j-edge-count]
{
  "type": "relationship",
  "id": "878",
  "label": "COMMUNICATES_TO",
  "start": {
    "id": "2891",
    "labels": [
      "IP1"
    ],
    "properties": {
      "ip_address": "192.168.1.100"
    }
  },
  "start": {
    "id": "2892",
    "labels": [
      "IP2"
    ],
    "properties": {
      "ip_address": "192.168.1.101"
    }
  },
    "properties": {
      "count": 2,    
      "id": "6ec4f95c-f4e3-4516-92c2-172cec275696"
    }
  }
\end{lstlisting}

\begin{lstlisting} [language=YARA, mathescape=true, caption={IRSKG GNN computational model input data schema representing transformed vertex, the router, $NEP_2$, with ip-address=10.10.10.10 based on $\RoE_1$ (shown in Listing \ref{list:network-case-study-ingestion-phase-dataingestion-neo4j-fw_rule} and in Figure \ref{fig:AICA-IRS-KG-NMS-ROE}) has a \textit{`count'} property value of -1000000 per Equation \ref{eq:gnn-sensor-graph-transform-edge-constraint}.}, label=list:network-case-study-ingestion-phase-datatransform-neo4j-vertex-roe-count]
{
  "type": "node",
  "id": "27550",
  "labels": [
    "NEP2"
  ],
  "properties": {
    "ip_address": "10.10.10.10",
    "count": -1000000
  }
}
\end{lstlisting}

\begin{figure}[ht]
    \centering
    \includegraphics[scale=0.42]{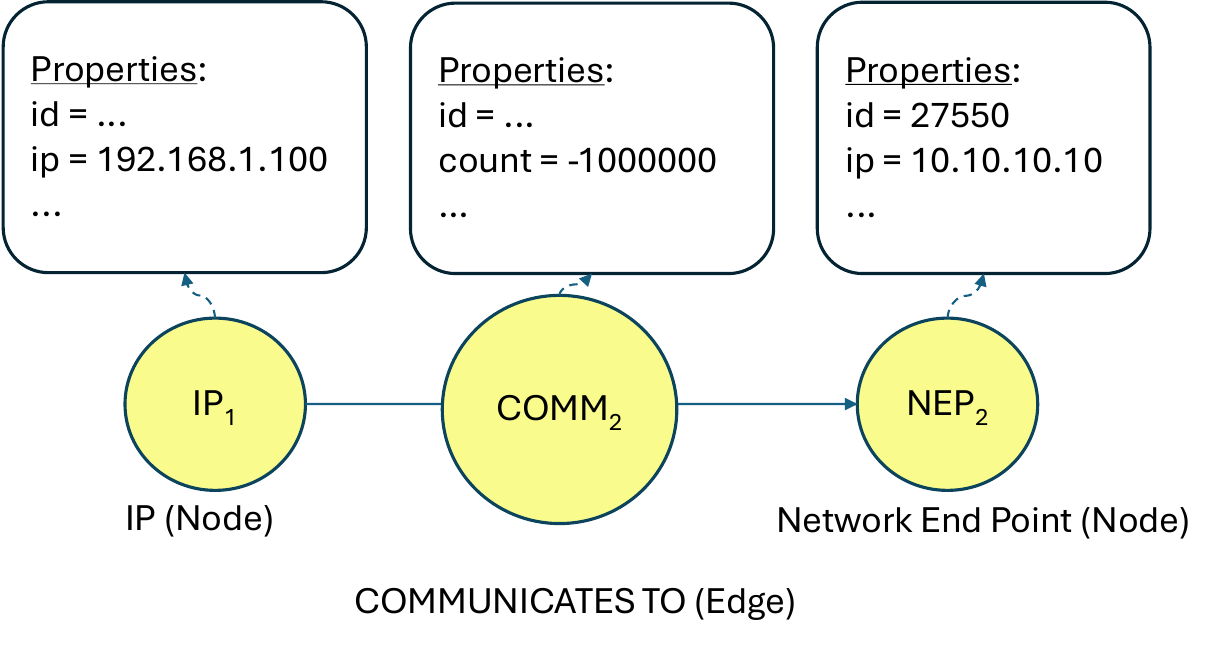}
    \caption{IRSKG GNN computational model input data schema using the Gray log and the constraint rule, that denies modifying the router. Graylog IRSKG is shown in the Fig. \ref{fig:AICA-IRS-KG-NMS-Senses} and its Neo4J implement is shown in the Listing \ref{list:network-case-study-ingestion-phase-dataingestion-raw-graylog}. The constraint IRSKG is shown in Fig. \ref{fig:AICA-IRS-KG-NMS-ROE} and its Neo4J implementation is shown in the Listing \ref{list:network-case-study-ingestion-phase-dataingestion-neo4j-fw_rule}. Based on $RoE$, we represent $\pr_{count}(E_{1,2})$ as $-1000000$ to the Edge, $\E_{1,2}=COMM_2$, between $P_1$ with $ip\_address=192.168.1.100$ and $NEP_2$ with $ip\_address=10.10.10.10$. }
    \label{fig:ICA-IRS-KG-NMS-GNN-Input}
\end{figure}

Finally, we transform the IRSKG Gray logs using the constraint IRSKG rules to the IRSKG GNN computational input IRSKG. As explained in Section \ref{rules-data-model}, an enterprise administrator creates a constraint rule, $\RoE_1$, to prevent any source IP from connecting to a critical infrastructure piece, the router, by adhering to the rules of engagement semantics as illustrated in the Equation \ref{eq:gnn-sensor-graph-transform-edge-constraint} and demonstrated in Listing \ref{list:network-case-study-ingestion-phase-dataingestion-neo4j-fw_rule} and in Figure \ref{fig:AICA-IRS-KG-NMS-ROE}. $\RoE_1$ prevents any machines from connecting to the router, thus denying them the ability to add new rules to the router. The constraint rule transforms the router vertex and the edge to vertex from any source IP and assigns the edge $count$ property to $-1000000$. The transformed GNN computational model input IRSKG schema is shown in Fig. \ref{fig:ICA-IRS-KG-NMS-GNN-Input}. The ISRKG Neo4J implementations for the Vertex and the Edge are shown in the Listing \ref{list:network-case-study-ingestion-phase-datatransform-neo4j-vertex-roe-count} and \ref{list:network-case-study-ingestion-phase-datatransform-neo4j-edge-roe-count}.

\begin{lstlisting} [language=YARA, mathescape=true, caption={IRSKG GNN computational model input data schema representing transformed edge (as shown in Figure \ref{fig:ICA-IRS-KG-NMS-GNN-Input}), $COMM_2$, between the router, $NEP_2$ with ip-address=10.10.10.10 and vertex, $IP_1$ with ip-address=192.168.1.100 based on $\RoE_1$ (shown in Listing \ref{list:network-case-study-ingestion-phase-dataingestion-neo4j-fw_rule} and in Figure \ref{fig:AICA-IRS-KG-NMS-ROE}) has a \textit{`count'} property value of -1000000 per Equation \ref{eq:gnn-sensor-graph-transform-edge-constraint}.}, label=list:network-case-study-ingestion-phase-datatransform-neo4j-edge-roe-count]
{
  "type": "relationship",
  "id": "878",
  "label": "COMM2",
  "start": {
    "id": "2891",
    "labels": [
      "IP1"
    ],
    "properties": {
      "ip_address": "192.168.1.100"
    }
  },
  "start": {
    "id": "27550",
    "labels": [
      "NEP2"
    ],
    "properties": {
      "ip_address": "10.10.10.10"
    }
  },
    "properties": {
      "count": -1000000,    
      "id": "6ec4f95c-f4e3-4516-92c2-172cec275696"
    }
  }
\end{lstlisting}  

\section{Conclusion}
\label{conclusion}
The goal of the paper is to introduce a novel schema, called IRSKG, for capturing information to enhance cyber defense Intrusion Response Systems (IRSs). The schema accomplishes this by enabling: faster onboarding of new enterprise systems, brisker IRS rules management, and faster input data transformations to continuously train computation models to thwart security breaches. Additionally, IRSKG is designed to be adaptable to the evolving cyber threat landscape and allows the onboarding of new configurable structures. This schema represents enterprise system information, including Enterprise system logs, IRS rules, computation model input data, and chat conversation history. We chose a Network Infrastructure Management system as a case study using the AICA Prototype software for the demonstration. Using IRSKG, we represented Graylog network logs, IRS rules that govern the network path between hosts, and the model input to train the GNN. We considered GNN for demonstration due to the graphical nature of the data structure; however, one could use any technique and AI/ML model type to implement IRSKG. Moreover, one could choose a different prototype and a case study to prototype IRSKG. This unified and robust approach allows streamlined automated intrusion response with collaborative information sharing and explainability. In the future, we plan to automate our approach further by incorporating a set of programming APIs and tools to provide additional methods to interact with the IRSKG schema-compliant data. We want to use the APIs to ingest enterprise systems' logs, IRS rules, and computation model input data to use the tools and manage the data by visualizing the schema.

\section{Acknowledgements}

The work was supported by PATENT Lab at the Department of Computer Science and Engineering, Mississippi State University.
The views and conclusions are those of the authors.

\bibliographystyle{IEEEtran}
\bibliography{aica-knowledge-graph}


\end{document}